\documentclass[submitting]{nst}
\usepackage{aas_macros}

\usepackage{graphicx}
\usepackage{amssymb,amsmath}
\usepackage{soul}
\usepackage{xspace}
\newcommand{\src}{B0540--69\xspace}
\newcommand{\xrt}{Swift/XRT\xspace}

\newcommand{\red}[1]{#1}
\makeatletter
\providecommand{\href}[2]{#2}
\providecommand{\href@noop}[2]{#2}
\providecommand{\Eprint}[2]{#2}
\providecommand{\doibase}{https://doi.org/}
\providecommand{\enquote}[1]{``#1''}
\makeatother

\usepackage{CJK}
\begin{document}
\begin{CJK*}{UTF8}{gbsn}

\title{Flux Evolution of PSR B0540--69 and the Spectral Energy Distribution of its Pulsar Wind Nebula}

% zhangj*, lixz, zhangjy, penghl, gemy*, chensz*, ,wengss, zhengsj, lick, jiasm, cheny

% 暂定 author list如下：
\author{Juan Zhang(张娟)}
\email[Corresponding author, ]{zhangjuan@ihep.ac.cn}
\affiliation{State Key Laboratory of Particle Astrophysics, Institute of High Energy Physics, Chinese Academy of Sciences, 100049 Beijing, China}
\affiliation{University of Chinese Academy of Sciences, 100049 Beijing, China}

\author{Xuezhi Liu(刘学智)}
\affiliation{Institute of Astrophysics, Central China Normal University, Wuhan 430079, China}
\affiliation{State Key Laboratory of Particle Astrophysics, Institute of High Energy Physics, Chinese Academy of Sciences, 100049 Beijing, China}

\author{Junyang Zhang(张俊洋)}
\affiliation{State Key Laboratory of Particle Astrophysics, Institute of High Energy Physics, Chinese Academy of Sciences, 100049 Beijing, China} 
\affiliation{University of Chinese Academy of Sciences, 100049 Beijing, China}
\affiliation{TIANFU Cosmic Ray Research Center, 610000 Chengdu, Sichuan, China}

\author{Hanlong Peng(彭寒龙)}
\affiliation{School of Physics and Technology, Nanjing Normal University, Nanjing, 210023, Jiangsu, China}
% \affiliation{State Key Laboratory of Particle Astrophysics, Institute of High Energy Physics, Chinese Academy of Sciences, 100049 Beijing, China}

\author{Mingyu Ge(葛明玉)}
\email[Corresponding author, ]{gemy@ihep.ac.cn}
\affiliation{State Key Laboratory of Particle Astrophysics, Institute of High Energy Physics, Chinese Academy of Sciences, 100049 Beijing, China}
\affiliation{University of Chinese Academy of Sciences, 100049 Beijing, China}

\author{Songzhan Chen(陈松战)}
\email[Corresponding author, ]{chensz@ihep.ac.cn}
\affiliation{State Key Laboratory of Particle Astrophysics, Institute of High Energy Physics, Chinese Academy of Sciences, 100049 Beijing, China} 
\affiliation{University of Chinese Academy of Sciences, 100049 Beijing, China}
\affiliation{TIANFU Cosmic Ray Research Center, 610000 Chengdu, Sichuan, China}

\author{Shanshan Weng(翁山杉)}
\affiliation{School of Physics and Technology, Nanjing Normal University, Nanjing, 210023, Jiangsu, China}

\author{Shijie Zheng(郑世界)}
\affiliation{State Key Laboratory of Particle Astrophysics, Institute of High Energy Physics, Chinese Academy of Sciences, 100049 Beijing, China}
\affiliation{University of Chinese Academy of Sciences, 100049 Beijing, China}

\author{Shumei Jia(贾淑梅)}
\affiliation{State Key Laboratory of Particle Astrophysics, Institute of High Energy Physics, Chinese Academy of Sciences, 100049 Beijing, China}
\affiliation{University of Chinese Academy of Sciences, 100049 Beijing, China}

\author{Chengkui Li(李承奎)}
\affiliation{State Key Laboratory of Particle Astrophysics, Institute of High Energy Physics, Chinese Academy of Sciences, 100049 Beijing, China}
\affiliation{University of Chinese Academy of Sciences, 100049 Beijing, China}

\author{Haisheng Zhao(赵海升)}
\affiliation{State Key Laboratory of Particle Astrophysics, Institute of High Energy Physics, Chinese Academy of Sciences, 100049 Beijing, China}
\affiliation{University of Chinese Academy of Sciences, 100049 Beijing, China}

\author{Ju Guan(关菊)}
\affiliation{State Key Laboratory of Particle Astrophysics, Institute of High Energy Physics, Chinese Academy of Sciences, 100049 Beijing, China}

\author{Yong Chen(陈勇)}
\affiliation{State Key Laboratory of Particle Astrophysics, Institute of High Energy Physics, Chinese Academy of Sciences, 100049 Beijing, China}
\affiliation{University of Chinese Academy of Sciences, 100049 Beijing, China}

\author{Liming Song(宋黎明)}
\affiliation{State Key Laboratory of Particle Astrophysics, Institute of High Energy Physics, Chinese Academy of Sciences, 100049 Beijing, China}
\affiliation{University of Chinese Academy of Sciences, 100049 Beijing, China}

\author{Fangjun Lu(卢方军)}
\affiliation{State Key Laboratory of Particle Astrophysics, Institute of High Energy Physics, Chinese Academy of Sciences, 100049 Beijing, China}
\affiliation{University of Chinese Academy of Sciences, 100049 Beijing, China}

\begin{abstract}
The young Crab-like pulsar PSR \src, located in the Large Magellanic Cloud, underwent a spin-down rate transition (SRT) in 2011 December, after which its pulsar wind nebula (PWN) was reported to brighten up to 2018. In this work, we extend the study of its X-ray and $\gamma$-ray flux evolution through 2026 by combining observations from multiple X-ray instruments with data from the \textit{Fermi} Large Area Telescope, thereby constructing a flux history of the PSR$+$PWN system spanning more than two decades. We find that the enhanced X-ray luminosity of the PSR$+$PWN system has persisted for more than 14~yr after the SRT, up to early 2026, whereas the pulsed X-ray flux remains unchanged across all epochs. This confirms that the brightening originates from the PWN. In the $\gamma$-ray band, neither the pulsed nor the non-pulsed flux shows any detectable long-term variation between 2008 and 2025. Together, these results imply that the additional spin-down energy released at the SRT has been continuously injected into the PWN over decades, placing strong constraints on the energy budget and physical mechanism of the transition. Combining the non-pulsed X-ray and $\gamma$-ray fluxes measured here with previously published radio-to-UV data, we further construct a broadband spectral energy distribution of the PWN, which shows properties markedly different from those of the Crab Nebula.
\end{abstract}


\maketitle

\nolinenumbers

\section{Introduction}
\label{sec:intro}

First discovered in soft X-rays by the Einstein Observatory \citep{1984ApJ...287L..19S}, the pulsar PSR~\src\ (also known as PSR~J0540$-$6919) has subsequently been detected across the electromagnetic spectrum from radio to $\gamma$-rays \citep{1993ApJ...403L..29M, 2004MNRAS.355...31J, 2011MNRAS.412.2689G, 2019ApJ...871..246M, 1999AA...348..519M, 2012ApJS..199...32G, 2015Sci...350..801F}. It is a 50\,ms pulsar with a characteristic age of $\sim1700$~yr and a spin-down power of $1.5\times10^{38}$\,erg\,s$^{-1}$. Embedded in the supernova remnant (SNR)~0540$-$693 in the Large Magellanic Cloud (LMC), at a precisely determined distance of $49.59 \pm 0.09\,\mathrm{(stat.)} \pm 0.54\,\mathrm{(sys.)}$~kpc \citep{2019Natur.567..200P}, it powers the luminous pulsar wind nebula (PWN) N158A resolved by \emph{Chandra} \citep{2000ApJ...532L.117G,2007ApJ...662..988P}. Owing to its similarities in period, spin-down power, and nebular properties, PSR~\src is widely regarded as the extragalactic ``twin'' of the Crab pulsar.

In 2011 December, PSR~\src underwent a sudden spin-down rate transition (SRT) \citep{2015ApJ...807L..27M}: the magnitude of the frequency derivative $|\dot{\nu}|$ jumped by $\sim36\%$, from a low spin-down state of $\dot{\nu}\sim1.9\times10^{-10}$\,Hz\,s$^{-1}$ to a high spin-down state of $\dot{\nu}\sim2.5\times10^{-10}$\,Hz\,s$^{-1}$, without any detectable change in the rotation frequency or in the pulsed X-ray emission. The braking index $n \equiv \nu\ddot{\nu}/\dot{\nu}^2$, which had been stable at $\sim2.1$ before the SRT \citep{2007ApSS.308..317L, 2012ApJS..199...32G, 2015ApJ...812...95F}, dropped to nearly zero and has since recovered almost monotonically, reaching $\sim1.6$ by 2025 \citep{2019NatAs...3.1122G, 2020MNRAS.494.1865W, 2024ApJ...973L..39E, 2026MNRAS.546f2264H}. This event offers a rare opportunity to observe, in real time, the response of a PWN to a sudden change in the energy output of its central pulsar. Proposed theoretical interpretations of the post-SRT evolution include magnetic-axis drift driven by crustal platelet dynamics \citep{2026ApJ...998..319B}, magnetospheric reconfiguration \citep{2020MNRAS.494.1865W}, and fallback-accretion-induced field growth \citep{2017MNRAS.469.1974E}, but the underlying physical origin remains debated. Continued monitoring of the timing behavior and flux evolution of \src is therefore crucial for constraining the braking mechanism, the coupling between the pulsar spin state and the PWN, and the emission processes and structure of the pulsar magnetosphere. %\footnote{In the main text, unless otherwise stated, ``PSR~\src'' refers to the pulsar itself, ``the \src PWN'' to the surrounding pulsar wind nebula, and ``\src'' (or equivalently the ``\src PSR$+$PWN'') to the combined pulsar-plus-nebula system. The title is an exception: there, ``Pulsar B0540--69'' is used as the conventional name of the whole source system.}
 
The most striking observational consequence of the SRT was the brightening of the PWN. \citet{2019NatAs...3.1122G} showed that the observed flux of the PSR$+$PWN system of \src rose from $\sim4\times10^{-11}$\,erg\,s$^{-1}$\,cm$^{-2}$ before the SRT to $\sim4.7\times10^{-11}$\,erg\,s$^{-1}$\,cm$^{-2}$ afterwards, while the pulsed X-ray emission and the photon index of the PSR$+$PWN system remained unchanged. This behavior implied that the brightening arose from the PWN itself and provided direct evidence that the additional spin-down energy was channeled into enhanced particle injection into the nebula, rather than into increased magnetospheric emission from the pulsar \citep{2019NatAs...3.1122G}. The elevated flux of the PSR$+$PWN system was found to persist at least until 2018.  

In the $\gamma$-ray band, \citet{2023MNRAS.520.4456N} identified week-timescale flaring activity in the light curves of PSR~J0540$-$6919 and PSR~J0537$-$6910 using \textit{Fermi} Large Area Telescope (\textit{Fermi}/LAT) data from 2008 to the end of 2021. However, no long-term $\gamma$-ray flux evolution or clear connection to the X-ray enhancement around the SRT was reported. The observed flares resemble the day-timescale flares of the Crab Nebula \citep{2011Sci...331..736T,2011Sci...331..739A}, showing similarly hard spectral indices between 100 and 300~MeV to that of the Crab Nebula giant flare \citep{2012ApJ...749...26B,2023MNRAS.520.4456N}. In contrast, the 1--10~GeV flares from the LMC pulsars display no counterparts at lower energies, suggesting that different particle-acceleration mechanisms may be at work \citep{2023MNRAS.520.4456N}. A transient high $\gamma$-ray state of PSR~J0540$-$6919 during 2025 December 22--29 was also reported by \citet{2026ATel17572....1M} using \textit{Fermi}/LAT data.
  
Very-high-energy (VHE, $E>100$~GeV) observations can probe the acceleration limit of the pulsar wind and constrain the radiation mechanism. Imaging Atmospheric Cherenkov Telescopes (IACTs) such as H.E.S.S., VERITAS, and MAGIC have resolved dozens of Galactic PWNe at TeV energies, tracing the highest-energy electrons through their inverse-Compton emission and providing crucial constraints on pulsar-wind models. More recently, the Large High Altitude Air Shower Observatory (LHAASO) has extended detections into the ultra-high-energy regime (UHE, $E>100$~TeV), detecting photons up to 1.1~PeV from the Crab Nebula and firmly establishing PWNe as the most luminous class of Galactic PeVatrons \citep{2021Natur.594...33C,2021Sci...373..425L,2024ApJS..271...25C}. Despite its close similarity to the Crab, \src and its associated PWN N158A have not yet been detected at VHE, owing to the large distance to the source and the current sensitivity of southern-hemisphere IACTs; H.E.S.S. observations have so far provided only an upper limit in the TeV band \citep{2015Sci...350..801F}. 

In this work we compile X-ray and $\gamma$-ray measurements from the Follow-up X-ray Telescope aboard the Einstein Probe (EP/FXT), the Swift X-ray Telescope (Swift/XRT), the Neutron star Interior Composition Explorer (NICER), and \textit{Fermi}/LAT, supplemented by archival X-ray data, to construct the most comprehensive flux history of the \src PSR$+$PWN system compiled to date, spanning from 2000 to early 2026. We attribute the off-pulse emission to the PWN and use it to build the spectral energy distribution (SED) of the nebula. Broadband SED modelling of PWNe is a powerful diagnostic of the underlying electron population and magnetic field, because different wavebands probe distinct portions of the synchrotron and inverse-Compton spectra \citep{2014JHEAp...1...31T}. A joint SED analysis therefore offers the most direct route to simultaneously constraining the electron spectrum, magnetic field, and evolutionary state of this young, Crab-like PWN. 

The paper is organized as follows. The observations and data reduction procedures for each instrument are described in Section~\ref{sec:obs}. The timing results derived from EP/FXT, NICER, and \textit{Fermi}/LAT data are presented in Section~\ref{sec:timing} and are used to resolve the on- and off-pulse spectra. The X-ray and $\gamma$-ray flux evolution is presented in Section~\ref{sec:flux}. Building on the off-pulse luminosity, the SED of the PWN and possible physical interpretations are explored in Section~\ref{sec:sed}. Our discussion and conclusions are given in Section~\ref{sec:dis}.

\section{Observation and Data Reduction}\label{sec:obs}

\subsection{EP/FXT}\label{sec:fxt}

The EP/FXT monitoring campaign for \src began in January 2026, with the first observing run concluding in February 2026. The campaign consisted of four observations covering approximately 50 days and was carried out in FXT partial-window (PW) mode. A log of the observations is given in Table~\ref{tab:obslog}. The data were reduced with \texttt{fxtchain} from the FXT Data Analysis Software (FXTDAS \citep{2025RDTM....9..215Z}) v1.30 and the FXT Calibration Database (CALDB \citep{2025RDTM....9..250L}) v1.30. Good time intervals (GTIs) were selected using the criteria \texttt{ELV>5 \&\& SAA==0 \&\& DYE\_ELV>30}, i.e., excluding intervals inside the South Atlantic Anomaly and requiring Earth (bright-Earth) elevation angles larger than $5^\circ$ ($30^\circ$). The \texttt{fxtbary} task was used to convert Terrestrial Time (TT) to Solar-system Barycentric Dynamical Time (TDB), adopting the JPL \texttt{DE405} ephemeris and the \texttt{SIMBAD} source position from \emph{Gaia} Early Data Release 3, R.A.~$=05^{\rm h}40^{\rm m}11.20^{\rm s}$, Dec.~$=-69^{\circ}19^{\prime}54.2^{\prime\prime}$ (J2000) \citep{2020yCat.1350....0G}. Source events were extracted from a circular region of $72^{\prime\prime}$ radius centered on this position, and background events were extracted from source-free box regions within the PW field of view. As the spectra from the four observations show no significant variation across the monitoring period, they were combined using the \texttt{addspec} tool. The total effective exposure obtained is 9.7\,ks for each FXT module.

\begin{table}[tb]
    \centering   
    \caption{Observation logs of the EP/FXT and Swift/XRT data of \src analyzed in this work. The exposure column lists the per-observation exposure time for each FXT module, and, for Swift/XRT, the combined exposure time grouped by observing mode and year.
    } 
    \label{tab:obslog}
    \begin{tabular}{ccccc}
    \hline
    Telescope & Date (YYYY-MM-DD) &  ObsID &   & Exposure (ks)  \\
              % &  YYYY-MM-DD    &        &   & ks     \\
    \hline
    EP/FXT    & 2026-01-03 & 08500000512 & &1.8 \\
              & 2026-01-16 & 08500000521 & &1.7 \\
              & 2026-01-30 & 08500000556 & &3.5 \\
              & 2026-02-20 & 08500000563 & &2.7 \\
    \hline
    Swift/XRT &  2018      & 00093144012 &  &       \\
              &            & 00093144013 &  &       \\
              &            & 00093144014 &  &       \\
              &            & 00033603055 &  &       \\
              &            & 00033603056 &  &       \\
              &            & 00033603057 &  &       \\
              &            & 00033603058 &  &       \\
              &            & 00033603059 &  & 28.1(WT)    \\
              &            & 00033603060 &  &   \\
              &            & 00033603061 &  &       \\
              &            & 00033603062 &  &       \\
              &            & 00033603063 &  &       \\
              &            & 00033603064 &  &       \\
              &            & 00033603065 &  &       \\
              &            & 00033603066 &  &       \\
              \cline{2-5}
              &  2019      & 00033603067 &  &  4.1(WT)     \\
              &            & 00033603068 &  &      \\
              % \cline{2-5}
              % &  2020      & 00045445002 &   &        \\
              \cline{2-5}
              &  2023      & 00015472001 &   & 1.4 (PC)       \\
              \cline{2-5}
              &  2026      & 00033603069 &   &        \\
              &            & 00033603070 &   &  1.9(WT)      \\
              &            & 00033603071 &   &  3.4(PC)      \\
              &            & 00033603072 &   &        \\
    % \hline
    %           & 2017-07-25     & 1020010101    &  57959    &  \\
    %  NICER    & to             &  to           &   to     & 80.6  \\
    %           & 2025-03-24     & 7508025601    &  60758 \\
    \hline
    \end{tabular}    
\end{table}

\subsection{Swift/XRT}\label{sec:xrt}

We use all available \xrt observations of \src taken in window-timing mode (WT) and photon-counting mode (PC) between 2018 and 2026, as listed in Table\,\ref{tab:obslog}. Spectra were generated with the online tool provided by the UK Swift Science Data Centre\footnote{\url{https://www.swift.ac.uk/user_objects/}} \citep{2009MNRAS.397.1177E} and grouped to a minimum of 25 counts per bin. No spectrum was extracted for 2020, since \src fell outside the field of view of the corresponding ToO observation; in the 2023 pointing the source was located $4.7^{\prime}$ off-axis. The effective exposure times obtained in each year, grouped by observing mode, are summarized in Table\,\ref{tab:obslog}.
% for this observation, the spec fit was conducted with fixed nH and photon index.
 %28.1\,ks (WT, 2018), 4.1\,ks (WT, 2019), 1.4\,ks (PC, 2023), 1.9\,ks (WT, 2026), and 3.4\,ks (PC, 2026).

\subsection{NICER}\label{sec:nicer}

NICER has monitored \src from 2017 July to 2025 June. In this paper we analyze all 212 NICER observations of \src available to date (ObsIDs 1020010101 to 7508025601), covering MJD 57959--60758. Data reduction is performed with \texttt{HEASOFT} v6.36. We first apply the standard NICER data-reduction pipeline: the \texttt{nicerl2} task (included in \texttt{NICERDAS} v15) is run with default parameters to produce cleaned and calibrated event files. To mitigate contamination from particle-induced flares, we extract 12--15\,keV light curves with \texttt{nicerl3-lc} and use \texttt{maketime} to reject intervals with count rates exceeding $0.2$\,counts\,s$^{-1}$. The resulting GTI files are then applied by re-running \texttt{nicerl2}, producing the final event files. Barycentric correction is performed with the \texttt{barycorr} tool in the \texttt{HEASOFT FTOOLS} package, adopting the \texttt{DE405} ephemeris. Finally, we use the \texttt{nicerl3-spect} tool to generate energy spectra and background products, adopting the \texttt{3C50} background model. For a more comprehensive description of the NICER data-processing pipeline, we refer the reader to the official NICER analysis webpage\footnote{\url{https://heasarc.gsfc.nasa.gov/docs/nicer/nicer_analysis.html}}.

After the data reduction, we retain only observations with exposure times longer than 50\,s for subsequent spectral analysis. The total exposure obtained is 80.6\,ks. We perform spectral fitting for each individual observation in \texttt{XSPEC} using a \texttt{powerlaw} model. As the key parameters (flux and photon index) show no significant variations across observations, we merge the spectra with the \texttt{addspec} tool to improve the signal-to-noise ratio. This procedure yields 13 high-quality combined spectra, with response files merged accordingly using \texttt{addrmf} and \texttt{addarf}. %The resulting spectra are then fitted with an absorbed power-law model (\texttt{cflux*tbabs*powerlaw}), and the fitting results are presented in Figure~\ref{fig:luminosity}.

\subsection{\textit{Fermi}/LAT}\label{sec:fermi}
We extracted \textit{Fermi}/LAT data from 2008 August 4 to 2025 June 9, selecting \texttt{P8R3\_SOURCE\_V3} events in the energy range 100\,MeV--300\,GeV within a $15^{\circ}\times 15^{\circ}$ region centered on B0540$-$69. The data were further filtered with the criteria $z<90^\circ$ (zenith angle), \texttt{DATA\_QUAL>0\,\&\&\,LAT\_CONFIG==1}, \texttt{evclass=128}, and \texttt{evtype=3}. The \texttt{gtbary} tool was used to apply the barycentric correction with the JPL DE405 Solar-system ephemeris. \red{We retained the full LAT data set in the main analysis, including the flare intervals reported by \citet{2023MNRAS.520.4456N}, and assessed separately whether their removal affects the inferred $\gamma$-ray fluxes and spectra (Section~\ref{sec:fermi-flux}).}

\section{Timing Analysis}\label{sec:timing}

To resolve the pulsed emission of \src, we combined a subset of the NICER data with the new timing data from FXT in order to extend the ephemeris coverage provided by \citet{2026MNRAS.546f2264H}. In contrast to the spectral fitting, and in order to maximize the significance of the pulsed signal during the timing analysis, we used Stingray \citep{2019ApJ...881...39H, 2024JOSS....9.7389B} to discard intervals in the NICER data whose average count rate within each 8\,s segment exceeded $0.2\ \mathrm{counts\,s^{-1}}$, and we selected photons in the 0.85--7.3\,keV energy range. For the FXT data the energy range was restricted to 0.9--12\,keV. We also attempted to perform a timing analysis using the LAT data; however, even after folding 400 days of data with the available ephemeris, the weighted pulse profile \citep{2019AA...622A.108B} reached only $\sim4\sigma$ significance. \red{This is insufficient to establish reliable timing solutions for the intervals not covered by the ephemerides of \citet{2012ApJS..199...32G,2015ApJ...807L..27M,2019NatAs...3.1122G} and \citet{2026MNRAS.546f2264H}.}

\begin{figure}[htb]
    \centering
    \includegraphics[width=\linewidth]{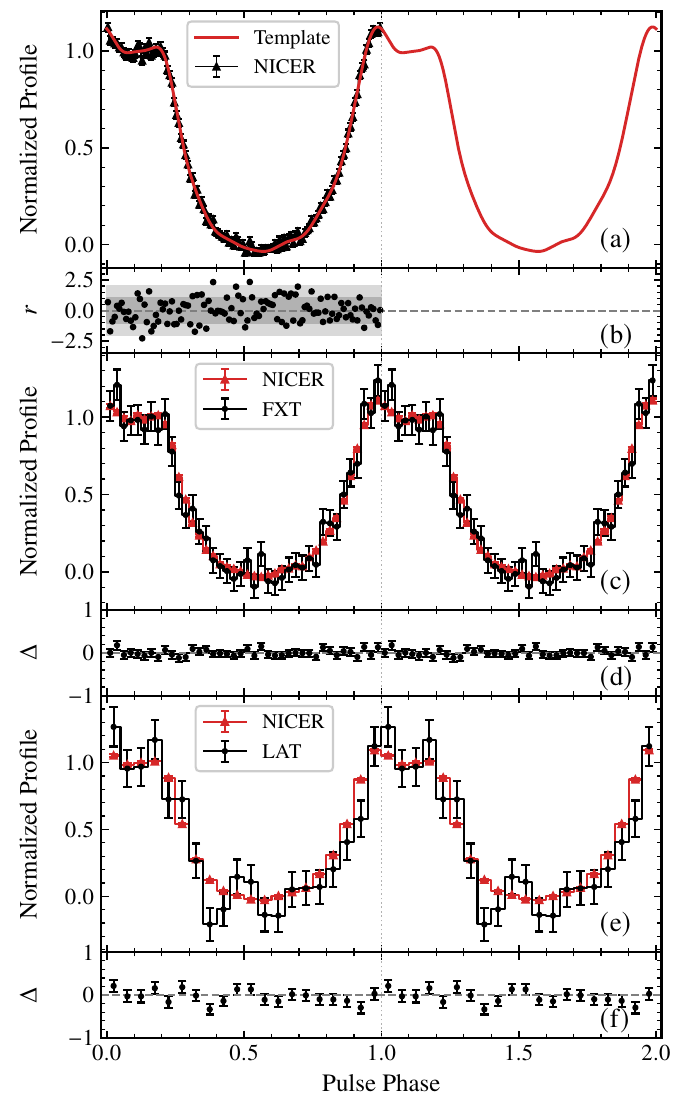}
    \caption{\red{Pulse-profile comparison among NICER, FXT, and LAT. Panel~(a) shows the normalized NICER profile with 120 phase bins and the continuous template constructed from a tenth-order Fourier series. The NICER data are shown only over phases 0--1, whereas the template is repeated over phases 0--2 so that only the template shape is displayed during the second cycle. Panel~(b) shows the bin-by-bin Pearson residuals over phases 0--1, defined as $r=(D-M)/\sqrt{M}$, where $D$ and $M$ denote the observed and model counts, respectively. Panels~(c) and (e) compare the normalized NICER profile with the FXT and LAT profiles using 40 and 20 phase bins, respectively; the profiles are repeated over two rotational cycles for clarity. Panels~(d) and (f) show the corresponding residuals $\Delta$ of the FXT and LAT profiles relative to the NICER profile. All error bars represent $1\sigma$ uncertainties.}}
    \label{fig:profile2}
\end{figure}

To construct a new ephemeris, we converted the photon arrival times into a set of pulse times of arrival (TOAs) with their statistical uncertainties, and subsequently updated the ephemeris by fitting these TOAs. Specifically, we first folded all NICER data to obtain a reference pulse profile using the truncated Taylor expansion
\begin{equation}
    \phi = \phi_0 + \nu(t-t_0) + \frac{1}{2}\dot{\nu}(t-t_0)^2 + \frac{1}{6}\ddot{\nu}(t-t_0)^3,
\end{equation}
where $t_0$ is the reference epoch, $\phi_0$ is the corresponding phase, and $\nu$, $\dot{\nu}$, and $\ddot{\nu}$ are the spin frequency and its first and second derivatives, respectively, taken from the existing ephemeris. \red{The resulting reference profile was modeled empirically with a tenth-order Fourier series, whose best-fit continuous form was adopted as the X-ray pulse-profile template of \src{}. The template and the NICER profile are shown in Figure~\ref{fig:profile2}(a). The Pearson residuals shown in Figure~\ref{fig:profile2}(b) exhibit no apparent systematic structure, and the fit yields a reduced chi-square of $\chi_\nu^2=1.08$, indicating that the template provides an adequate description of the pulse profile.}

Next, we divided the NICER data spanning MJD 60628--60838 into 20 segments, \red{each containing approximately the same number of photons, and grouped the FXT data spanning MJD 61043--61092} into 8 segments according to observation ID and detector. Each of these 28 segments was folded using the final segment of the ephemeris of \citet{2026MNRAS.546f2264H}, yielding 28 profiles. We then performed an unbinned likelihood analysis \citep{2011ApJS..194...17R} between these profiles and the template to derive the phase differences and thereby obtain 28 TOAs with uncertainties. These TOAs were subsequently fitted with Tempo2 \citep{2006MNRAS.369..655H} to update the ephemeris. The procedure was iterated several times to refine the TOAs, yielding the final ephemeris listed in Table~\ref{tab:ephem}. \red{This latest ephemeris gives a braking index of $\sim1.53$, consistent with the trend of its long-term increasing evolution after the SRT \citep{2024ApJ...973L..39E, 2026MNRAS.546f2264H}. It is worth noting that this value has not yet reached the pre-SRT level of $\sim2.1$.}

\begin{table}[htb]
\caption{Timing parameters of PSR \src derived from NICER and EP/FXT observations. The source position, R.A.~$=05^{\rm h}40^{\rm m}11.20^{\rm s}$, Dec.~$=-69^{\circ}19^{\prime}54.2^{\prime\prime}$, is taken from SIMBAD \citep{2020yCat.1350....0G}.}
    \centering
    \begin{tabular}{lc}
    \hline
    Parameters     &   Values                       \\        
    \hline
    % R.A.           & $05^{\rm h}40^{\rm m}11\fs20$  \\
    % Dec.           & $-69\degr19\arcmin54\farcs2$   \\
    $\mathrm{Epoch}$ ($\mathrm{MJD}$)    &  60730                         \\
    $t_0$ ($\mathrm{MJD}$)    &  60804.80962990        \\
    $\nu$ ($\mathrm{Hz}$)         & 19.62108511955(90)          \\
    $\dot{\nu}$ ($10^{-10}\, \mathrm{Hz\,s}^{-1}$)    & -2.519086(20)     \\    
    $\ddot{\nu}$ ($10^{-21}\,\mathrm{Hz\,s}^{-2}$)    & 4.953(22)    \\     
    $\mathrm{MJD}$ range ($\mathrm{MJD}$)     & 60628--61092                   \\
    $\mathrm{RMS}$ ($\mu\mathrm{s}$ )           &  827                   \\
    $\chi^2/\mathrm{dof}$  & 38.90/24 \\
    \hline
    \end{tabular}    
    \label{tab:ephem}
\end{table}

\red{Using the new ephemeris together with the previously published timing solutions, Figure~\ref{fig:profile2} shows the NICER-derived pulse-profile template and compares the pulse profiles obtained with NICER, FXT, and LAT. For display, all profiles were linearly normalized by adopting the weighted mean over phases $0.4$--$0.6$ as the minimum level and that over the combined phase intervals $0$--$0.2$ and $0.9$--$1.0$ as the maximum level. Unlike in the X-ray band, the LAT profile was constructed using a photon-weighting method applied to events within a $3^\circ$-radius circular region centered on \src{}, thereby improving the pulse significance \citep{2019AA...622A.108B}.} In panel~(c) of Figure~\ref{fig:profile2}, the FXT profile almost overlaps the NICER profile, and both display a single broad pulse. In panel~(e), the $\gamma$-ray profile from LAT is also broadly consistent with the X-ray profile; the small bump around phase 0.4--0.6 is most likely due to statistical fluctuations rather than a genuine signal. The $\gamma$-ray profile derived here is also consistent with that reported by the \emph{Fermi}/LAT Collaboration \citep{2015Sci...350..801F}.

\section{Flux Evolution}\label{sec:flux}

\subsection{X-ray}

\begin{figure}[tb]
    \centering
    \includegraphics[width= \linewidth]{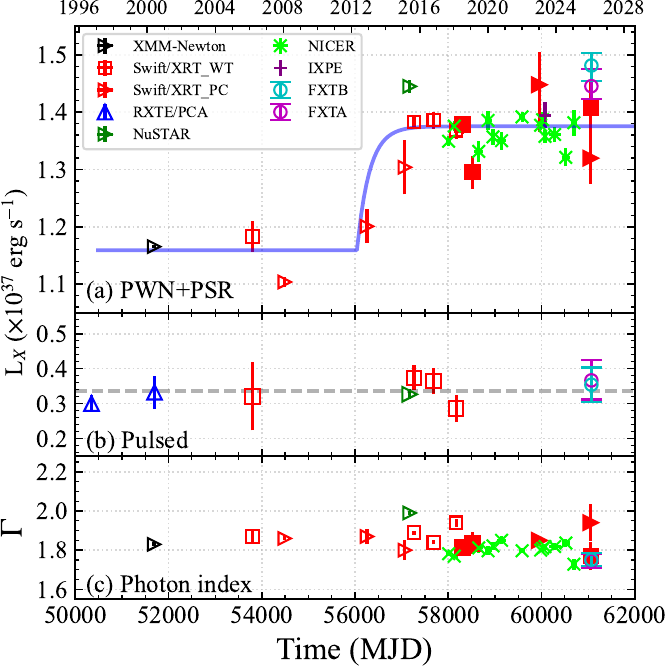}
    \caption{Evolution of the X-ray luminosity and photon index of \src. 
    FXTA and FXTB results are shown as magenta and cyan open circles, respectively. The other instruments analyzed in this work are NICER (green crosses) and Swift/XRT in PC mode (filled triangles) and WT mode (filled squares). 
    Hollow markers before 2018 are taken from Tables~1 and 2 of \citet{2019NatAs...3.1122G}. 
    (a)~Total 1--10~keV luminosity of the PSR$+$PWN system; the blue curve shows the best-fit model (Eq.~\ref{eq:lxmodel}). (b)~Pulsed luminosity; the horizontal dashed line marks its mean value. (c)~Photon index of the PSR$+$PWN system.}
    \label{fig:luminosity}
\end{figure}

The spectra of \src obtained from the EP/FXT, \xrt (2018--2026), and NICER (2017--2025) observations were fitted with an absorbed power-law model (\texttt{tbabs$\times$powerlaw}). The absorbed 1--10~keV fluxes were converted to luminosities via $L_X = 4\pi d^2 F_X$, adopting $d = 49.59$~kpc \citep{2019Natur.567..200P}. To trace the flux history of the PSR$+$PWN system, the resulting luminosities are shown in Figure\,\ref{fig:luminosity}(a), together with the archival results of \citet{2019NatAs...3.1122G}, which include the X-ray Multi-Mirror Mission (XMM-Newton), the Nuclear Spectroscopic Telescope Array (NuSTAR), the Rossi X-ray Timing Explorer Proportional Counter Array (RXTE/PCA), and earlier \xrt observations, as well as the Imaging X-ray Polarimetry Explorer (IXPE) measurement from 2023 \citep{2024ApJ...962...92X}. The flux evolution clearly shows that, before the SRT (2011 Dec 3, MJD~55898), the PSR$+$PWN system was stable at $L_X \approx 1.16 \times 10^{37}$~erg~s$^{-1}$. After the SRT, the luminosity rose to a plateau of $L_X \approx 1.38 \times 10^{37}$~erg~s$^{-1}$. The post-SRT luminosity varies within about 15\% but exhibits no secular trend of the kind observed in the Crab Nebula flux variations, which may be linked to changes in magnetic-field strength or to magnetosonic waves \citep{2011ApJ...727L..40W}. Thus the elevated post-SRT luminosity has persisted for more than 14~yr without any obvious systematic evolution. Figure\,\ref{fig:luminosity}(c) shows the photon index $\Gamma$, which remains in the range 1.8--1.9 throughout, without any  associated spectral evolution.

Figure\,\ref{fig:luminosity}(b) shows the pulsed luminosity. For the EP/FXT data, phase-resolved spectra were extracted using the pulse-phase information derived in Section~\ref{sec:timing}, and the pulsed flux was computed as $F_{\rm pulsed} = F_{\rm total} - F_{\rm off}$\footnote{\red{Since the off pulse phase may also have contributions from the pulsar magnetosphere, the obtained pulsed flux here should be interpreted as a lower limit.}}, adopting an off-pulse phase interval of $0.4$--$0.6$ and on-pulse intervals of $0$--$0.4$ and $0.6$--$1$ \red{(please refer to Appendix\,\ref{apx:onoffspc} for the on- and off-pulse spectra)}. The new EP/FXT measurements from 2026, when compared with the pre-SRT RXTE/PCA results, demonstrate that the pulsed flux remains constant at $L_{\rm pulsed} \approx 0.34 \times 10^{37}$~erg~s$^{-1}$.

Following \citet{2019NatAs...3.1122G}, we adopt the same model
\begin{equation}\label{eq:lxmodel}
    L_X(t) = L_{X,0}\left[1 + \epsilon\, H(t - t_1)\left(1 - e^{-(t-t_1)/\tau}\right)\right]
\end{equation}
to fit the luminosity evolution of the PSR$+$PWN system. Here $L_{X,0}$ is the pre-SRT baseline luminosity, $\epsilon$ the fractional increase, $\tau$ the e-folding timescale, and $t_1 = t_{\rm SRT} + t_d$ the onset of the PWN brightening, with $t_{\rm SRT}=55898$ and $t_d = 143$~d held fixed at the value derived from the termination-shock radius \citep{2007ApJ...662..988P}; $H$ denotes the Heaviside step function. The best-fit parameters, listed in Table\,\ref{tab:flux_fit}, confirm a step-like transition with $\tau \sim 268\pm91$~d.
\red{Using $\tau \sim 446.6 \rm{B}^{-3/2}\rm{E}^{-1/2}$\,d with weighted mean photon energy $\rm{E}=2.66$\,keV \citep{2019NatAs...3.1122G}, the corresponding inferred magnetic-field strength is  $1.01^{+0.32}_{-0.18}$\,mG.}
The post-SRT increase in the PWN luminosity is thus confirmed to be $26.3\pm0.5\%$. 

\begin{table}[htb]
\caption{Best-fit parameters of the PSR$+$PWN luminosity-evolution model.}
    \centering
    \begin{tabular}{lc}
    \hline
    Parameter & Value \\
    \hline
    $L_{X,0}$ ($\times10^{37}$\,erg\,s$^{-1}$) & $1.159 \pm 0.003$ \\
    $\epsilon$ & $0.187 \pm 0.004$ \\
    $\tau$ (days) & $268 \pm 91$ \\
    $t_{\rm SRT}$ (MJD) & 55898 (fixed) \\
    $t_d$ (days) & 143 (fixed) \\
    \hline
    \end{tabular}
    \label{tab:flux_fit}
\end{table}

\subsection{$\gamma$-ray}\label{sec:fermi-flux}
\red{Motivated by the clear post-SRT brightening of the off-pulse X-ray emission, we searched for an analogous response in the $\gamma$-ray band, considering the phase-averaged, pulsed, and non-pulsed components separately.} Using the ephemerides valid before MJD 57000 \citep{2012ApJS..199...32G,2015ApJ...807L..27M,2019NatAs...3.1122G} and thereafter \citep{2026MNRAS.546f2264H}, we divided the \textit{Fermi}/LAT data into 12 \red{phase-averaged and 10 phase-resolved intervals, each spanning approximately 500 days. For the phase-resolved analysis, the on-pulse intervals were defined as phases $0.0$--$0.3$ and $0.8$--$1.0$. Two longer data sets, MJD 54682--60835 and MJD 57044--59678, were used to construct the phase-averaged and phase-resolved reference spectral models, respectively.}

\red{We performed a binned likelihood analysis with \textit{fermipy} v1.4.3 \citep{2017ICRC...35..824W} and \textit{Fermitools} v2.5.3, using $0.1^\circ$ spatial bins and 35 logarithmic energy bins. The model comprised all 4FGL-DR4 sources within $30^\circ\times30^\circ$ of \src \citep{2022ApJS..260...53A,2023arXiv230712546B} plus the Galactic and isotropic diffuse emission. The phase-averaged spectrum of \src was modeled with \texttt{PLSuperExpCutoff4}; in the phase-resolved analysis, a full-phase \texttt{PowerLaw} described the non-pulsed emission and an on-pulse \texttt{PLSuperExpCutoff4} the pulsed emission.}

\red{All data sets were first processed with the same initial procedure: we ran \textit{fermipy} \texttt{optimize} and then fitted the normalizations of \src, the diffuse components, the four extended LMC components, and selected nearby or variable field sources.\footnote{\red{The latter included catalog sources with $\mathrm{TS}>20$ within $5^\circ$ of \src and variable sources within $8^\circ$.}} The resulting normalization correlations with \src were recorded for subsequent parameter selection.}

\begin{figure}[htb]
    \centering
    \includegraphics[width=\linewidth]{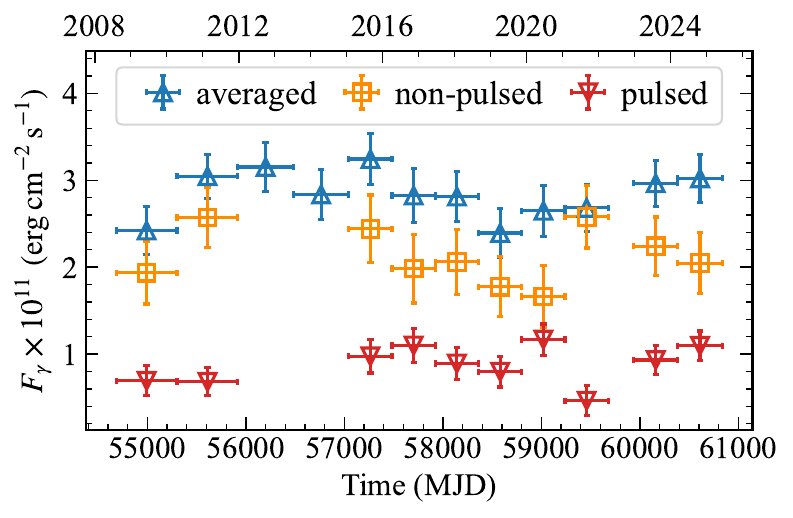}
    \caption{$\gamma$-ray flux evolution of B0540$-$69. The phase information for the pulsed (red) component was computed from the ephemerides of \citet{2012ApJS..199...32G, 2015ApJ...807L..27M, 2019NatAs...3.1122G} before MJD~57000 and of \citet{2026MNRAS.546f2264H} thereafter.}
    \label{fig:flux_fermi}
\end{figure}

\red{The two long-duration data sets were then used to determine the reference spectral shapes. Starting from their optimized models, we allowed the relevant \src components and only sufficiently correlated field sources to vary, fixing all other field sources. For each selected field source, all spectral parameters, including the normalization, were fitted.\footnote{\red{Field sources were selected when their normalization correlation satisfied $|\rho|>0.1$ with the phase-averaged component of \src, or with either the pulsed or non-pulsed component in the phase-resolved analysis.}} For each $\sim500$-day interval, the corresponding reference spectral shapes were fixed, and only the normalizations of the relevant \src components and correlated field sources were refitted.\footnote{\red{Field-source normalizations were refitted when the correlation measured in the initial fit of that interval satisfied $|\rho|>0.05$ with any relevant \src component.}} Thus, the long-duration data determined the reference spectral shapes, while the individual intervals measured the time-dependent normalizations.} The resulting flux evolution of \src is shown in Figure~\ref{fig:flux_fermi}.

\red{The phase-averaged, non-pulsed, and pulsed light curves show broadly similar temporal behavior, with the phase-averaged flux generally consistent with the sum of the two phase-resolved components. The third-to-last bin is an exception, with an unusually low pulsed flux and correspondingly high non-pulsed flux; because the two \src components are spatially coincident and the nearby PSR J0537$-$6910 introduces additional source confusion, we caution that this bin may be affected by imperfect flux partitioning, analogous to the ``flux transfer'' discussed in the 4FGL catalog \citep{2020ApJS..247...33A}. As an additional robustness check, we repeated the analysis after excluding the flare intervals reported by \citet{2023MNRAS.520.4456N}. The resulting fluxes changed by less than $0.5\sigma$, without materially affecting the light curves or subsequent conclusions. We therefore retained the full LAT data set in the main analysis.}

\red{To quantify whether any of the three $\gamma$-ray components underwent a step-like change at the SRT, we fitted their pre- and post-transition fluxes with a free ratio $R=F_{\rm post}/F_{\rm pre}$ and tested the fixed hypotheses $R=1$ and $R=1.3$ using $\Delta\chi^2$ with one degree of freedom. The latter provides a direct test of an increase comparable in magnitude to that seen in the X-ray band. The best-fit ratios are $1.028\pm0.076$, $1.337\pm0.248$, and $0.922\pm0.116$ for the phase-averaged, pulsed, and non-pulsed components, respectively. None shows a statistically significant departure from $R=1$ ($p=0.71$, $0.085$, and $0.53$). The pulsed component alone exhibits a nominal $\sim34\%$ increase, but remains consistent with both no change and a 30\% increase. By contrast, a 30\% increase is disfavored for the phase-averaged and non-pulsed components ($p=0.0034$ and $0.0139$), while remaining fully consistent with the pulsed component ($p=0.88$).}

\red{Because the only indication of an increase appears in the pulsed component rather than in the non-pulsed component, which provides the closest phase-resolved comparison with the off-pulse X-ray emission, we further tested whether the relative partition of the $\gamma$-ray flux between the two components changed across the SRT. Defining $Q=F_{\rm non\mbox{-}pulsed}/F_{\rm pulsed}$, we propagated the uncertainty in $\ln Q$ including the covariance between the two components obtained from the joint fit. The best-fit post/pre ratio is $R_Q=Q_{\rm post}/Q_{\rm pre}=0.657$, corresponding to a nominal $\sim34\%$ decrease, but the step model is not significantly preferred over a constant ratio ($\Delta\chi^2=2.37$, $p=0.124$).\footnote{\red{Because the third-to-last bin may be affected by imperfect flux partitioning, we repeated the test excluding this bin. The result remains similar, with $R_Q=0.618$, $\Delta\chi^2=3.07$, and $p=0.080$.}} Taken together, the flux and flux-ratio tests provide no statistically significant evidence for a $\gamma$-ray counterpart to the clear off-pulse X-ray brightening across the SRT. The nominal enhancement of the pulsed component remains suggestive, but the present data do not distinguish it from statistical fluctuations.}

\section{Spectral Energy Distribution and Modeling of \src PWN} \label{sec:sed}
Spatially resolved \emph{Chandra} observations of \src show that the PWN emission dominates over that of the surrounding SNR shell \citep{2007ApJ...662..988P}, as is commonly seen in young, energetic pulsar systems \citep{2006ARAA..44...17G}. The optical/near-infrared spectrum of \src further supports a non-thermal, most likely synchrotron, origin for its emission \citep{2019ApJ...871..246M,2024ApJ...966..125T}. \red{We therefore regard the off-pulse emission measured with EP/FXT and the non-pulsed emission measured with \textit{Fermi}/LAT as being dominated by the PWN.}

\red{To obtain the LAT measurements used in the broadband SED, we generated SEDs of the non-pulsed and pulsed component for two epochs using the \textit{fermipy} \texttt{sed} method, with the default configuration except for the energy binning. The 35 analysis energy bins were regrouped into nine SED bins: bins 1--32 were combined in groups of four, while bins 33--35 were combined into the final bin. For the later epoch, \texttt{sed} was applied directly to the best-fit phase-resolved reference model obtained over MJD 57044--59678, as described in Section~\ref{sec:flux}. For the earlier epoch, an additional data set spanning MJD 54682--55898 was processed using the same phase-resolved optimization and fitting procedure before applying \texttt{sed}. The resulting measurements are listed in Table~\ref{tab:sed_offpulse}.}

\begin{figure}[htb]
    \centering
    \includegraphics[width=\linewidth]{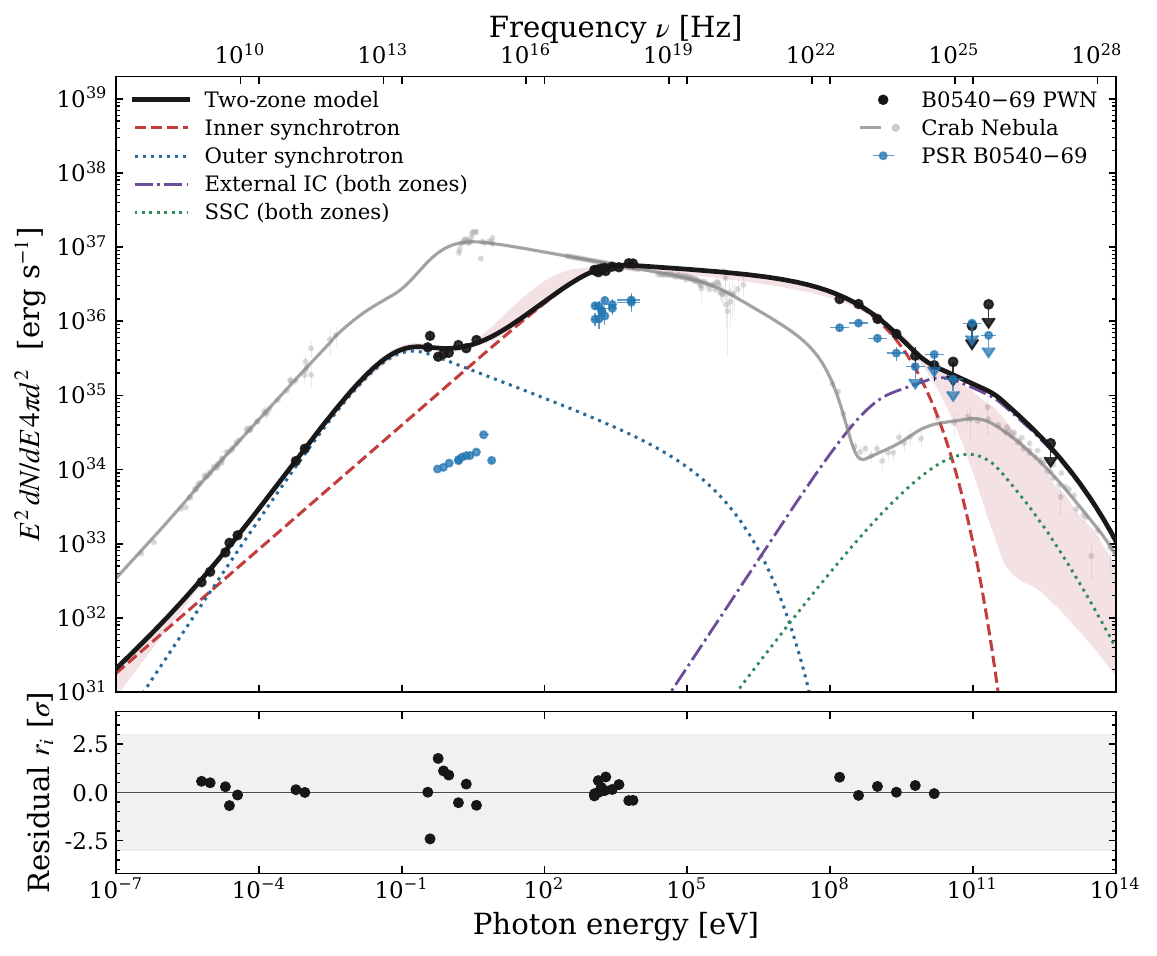}
\caption{\red{Upper panel: Broadband SED of the \src\ PWN (black points, with 95\% upper limits), overplotted with the best-fit two-zone model (black solid line; the shaded red band marks the 68\% conditional posterior predictive interval). For comparison, the Crab Nebula SED and its reference model (B = 124\,$\mu$G, d = 2\,kpc, and a SSC spatial scale of 2\,pc) from \citet{2010AA...523A...2M} are shown in grey. The phase-averaged SED of PSR B0540--69 itself (blue points) is also included: the NIR, optical and UV data are taken from \citet{2019ApJ...871..246M} and references therein, in which the reported flux corresponds to the pulsar itself and can therefore be regarded as an upper limit on the pulsed component; the FXT and LAT data are derived from the present analysis and represent the phase-averaged flux of the pulsed component. Lower panel: Normalized residuals of the fit.}}    
    \label{fig:b0540_two_zone_sed}
\end{figure}

We revisit the broadband SED of the PWN associated with \src\footnote{\red{The assembled data set should not be regarded as a strictly contemporaneous post-SRT SED. Apart from the Atacama Compact Array observations, most of the low-energy measurements were obtained before the SRT, whereas the \textit{Fermi}/LAT, EP/FXT, and H.E.S.S. data cover different observing epochs. We therefore make the relatively strong assumption that the low-energy synchrotron component did not vary significantly after the SRT.}} as shown in Figure~\ref{fig:b0540_two_zone_sed}. Also shown are the radio-to-UV measurements \citep{2020MNRAS.496.1834L}, the H.E.S.S.\ 99\% confidence-level upper limit in the direction of \src \citep{2015Sci...347..406H,2015Sci...350..801F}, and, for comparison, the SED of the Crab Nebula \citep{2010AA...523A...2M}, since both objects are powered by young pulsars with high spin-down luminosities and are prototypical young, luminous PWNe \citep{2002MNRAS.333..603H}.
Despite the close analogy with the Crab system, the \src\ PWN exhibits a markedly larger GeV-to-keV luminosity ratio, roughly an order of magnitude higher than that of the Crab Nebula \citep{2010AA...523A...2M,2023AA...671A..67D,2024AA...686A.308A}. 
%In addition, within the \textit{Fermi}/LAT band the SED of \src shows no clear transition region between the synchrotron and inverse-Compton components, in stark contrast to the well-defined structure observed in the Crab Nebula.
\red{In addition, the currently available SED of B0540--69 PWN is not yet sufficient to reveal a clear signature of the synchrotron-to-inverse-Compton (IC) transition, in contrast to the Crab Nebula, whose well-defined GeV spectral structure is attributed to a prominent synchrotron self-Compton (SSC) component. We note, however, for the Crab Nebula, the substantial SSC contribution makes it an atypical young PWN rather than a template for B0540--69 \citep{2010AA...523A...2M,2013MNRAS.436.3112T}.} 
Although the limited photometric coverage and the substantial gaps in the broadband SED preclude a precise determination of the luminosity peak, the available data indicate that this peak is unlikely to lie in the same energy band as that of the Crab Nebula.

\begin{table}[htb]
\centering
\caption{Off-pulse spectral energy distribution of \src. 
Fluxes are the unabsorbed $E^{2}\,dN/dE$; upper limits are 95\% C.L.}
\label{tab:sed_offpulse}
\begin{tabular}{lccc}
\hline\hline
Instrument / Band & $E$ & $E_{\rm lo}$--$E_{\rm hi}$ & $E^{2}\,dN/dE$ \\
                  &     &                             & ($10^{-12}$~erg~s$^{-1}$~cm$^{-2}$) \\
\hline
\multicolumn{4}{l}{\textit{EP/FXT-A} (keV)} \\
 & 1.128 & 1.020--1.235 & $16.5\pm0.81$ \\
 & 1.348 & 1.235--1.461 & $15.5\pm0.77$ \\
 & 1.589 & 1.461--1.717 & $17.4\pm0.87$ \\
 & 1.933 & 1.717--2.149 & $16.2\pm0.80$ \\
 & 3.660 & 2.149--5.170 & $18.2\pm0.88$ \\
 & 7.230 & 5.170--9.290 & $20.4\pm2.9$ \\
\multicolumn{4}{l}{\textit{EP/FXT-B} (keV)} \\
 & 1.116 & 1.013--1.219 & $16.7\pm0.82$ \\
 & 1.324 & 1.219--1.429 & $17.0\pm0.83$ \\
 & 1.556 & 1.429--1.684 & $16.9\pm0.82$ \\
 & 1.849 & 1.684--2.014 & $17.8\pm0.86$ \\
 & 2.642 & 2.014--3.270 & $18.5\pm0.89$ \\
 & 5.882 & 3.270--8.494 & $20.6\pm1.3$ \\
\multicolumn{4}{l}{\textit{Fermi-LAT (before SRT)} (GeV)} \\
 & 0.158 & 0.100--0.250 & \red{$7.80^{+0.80}_{-0.79}$} \\
 & 0.395 & 0.250--0.623 & \red{$5.17^{+0.65}_{-0.64}$} \\
 & 0.985 & 0.623--1.557 & \red{$2.65^{+0.55}_{-0.53}$} \\
 & 2.460 & 1.557--3.886 & \red{$2.46^{+0.60}_{-0.55}$} \\
 & 6.141 & 3.886--9.704 & \red{$1.13^{+0.65}_{-0.57}$} \\
 & 15.33 & 9.704--24.23 & \red{$<1.05$} \\
 & 38.28 & 24.23--60.49 & \red{$<0.83$} \\
 & 95.59 & 60.49--151.0 & \red{$<3.00$} \\
 & \red{212.9} & \red{151.0--300.0} & \red{$<5.40$} \\
\multicolumn{4}{l}{\textit{Fermi-LAT (after SRT)} (GeV)} \\
 & 0.158 & 0.100--0.250 & \red{$6.75^{+0.49}_{-0.49}$} \\
 & 0.395 & 0.250--0.623 & \red{$5.80^{+0.41}_{-0.40}$} \\
 & 0.985 & 0.623--1.557 & \red{$3.65^{+0.36}_{-0.35}$} \\
 & 2.460 & 1.557--3.886 & \red{$2.29^{+0.36}_{-0.34}$} \\
 & 6.141 & 3.886--9.704 & \red{$1.16^{+0.35}_{-0.31}$} \\
 & 15.33 & 9.704--24.23 & \red{$0.87^{+0.41}_{-0.34}$} \\
 & 38.28 & 24.23--60.49 & \red{$<0.96$} \\
 & 95.59 & 60.49--151.0 & \red{$<2.95$} \\
 & \red{212.9} & \red{151.0--300.0} & \red{$<5.75$} \\
\hline
\end{tabular}
\end{table}

To model the broadband SED of the \src\ PWN, \red{we calculate synchrotron, external IC, and local SSC emission using \texttt{naima}} \citep{2015ICRC...34..922Z}. We first tested a conventional one-zone model with a single electron population and a one-zone model with two electron populations. The former cannot adequately reproduce the multiwavelength SED, whereas the latter yields a better statistical fit. Nevertheless, the two-population one-zone model shows a pronounced degeneracy between the magnetic-field strength and the electron energy content, and requires rather extreme properties for the highest-energy electrons. We therefore do not adopt it as our preferred physical interpretation. There is clear observational evidence that the B0540$-$69 PWN is spatially inhomogeneous, as indicated by its resolved X-ray, radio, optical--near-infrared, and X-ray-polarimetric properties \citep{2000ApJ...532L.117G,2007ApJ...662..988P,2020MNRAS.496.1834L,2024ApJ...966..125T,2024ApJ...962...92X}. However, the currently available spatially resolved and multiwavelength information is insufficient to justify a highly parameterized spatio-temporal evolution model. We therefore construct  \red{a minimal spatially stratified effective two-zone model tied to the observed characteristic scales}.

The brightening of the X-ray PWN on a timescale of a few hundred days following the spin-down-rate transition of PSR~B0540$-$69 provides strong physical motivation for such a two-zone description \citep{2015ApJ...807L..27M,2019NatAs...3.1122G,2020MNRAS.494.1865W}. We distinguish a compact, X-ray-bright synchrotron nebula from a more extended, radio-emitting region. The inner zone is assumed to host a relatively strong magnetic field: \red{if the post-SRT X-ray response timescale is primarily set by synchrotron cooling, the estimate derived above is $B_{\rm in}=1.01^{+0.32}_{-0.18}\,\mathrm{mG}$. We fix $B_{\rm in}=1.01\,\mathrm{mG}$ in the model; it is an effective inner-zone field rather than a measurement of a uniform field throughout the PWN.} We confine this component to a compact region with a characteristic size comparable to that of the pulsar-wind termination shock. \red{The inner zone may contain high-energy electrons supplied by the pulsar wind and accelerated at or near the termination shock;} their synchrotron emission accounts for the bright X-ray radiation. The outer zone has a substantially weaker magnetic field and may contain a long-lived population of accumulated, lower-energy relic electrons. Their synchrotron emission contributes a large fraction of the radio-to-optical radiation, while their IC emission becomes increasingly important across the GeV--TeV band. In each zone, the electron distribution is described by an exponentially cut-off broken power law,
\begin{equation}
\begin{aligned}
N_i(E) ={}& A_i
\exp\left(-\frac{E}{E_{i,c}}\right) \\
&\times
\begin{cases}
\left(\dfrac{E}{E_0}\right)^{-p_{i,1}},
& E<E_{i,b}, \\[1.5mm]
\left(\dfrac{E_{i,b}}{E_0}\right)^{p_{i,2}-p_{i,1}}
\left(\dfrac{E}{E_0}\right)^{-p_{i,2}},
& E\geq E_{i,b},
\end{cases}
\end{aligned}
\label{eq:two_zone_electron_distribution}
\end{equation}
where $i\in\{\mathrm{in},\mathrm{out}\}$ and $E_0=1\,\mathrm{TeV}$. Here $p_{i,1}$ and $p_{i,2}$ denote the electron spectral indices below and above the break, $E_{i,b}$ is the break energy, and $E_{i,c}$ is the exponential cutoff energy. \red{The  photon fields for IC scattering comprise the CMB, an infrared component ($T=40\,\mathrm{K}$ \citep{2011ApJ...741...40T}), and an optical component ($T=4000\,\mathrm{K}$ \citep{2011ApJ...741...40T}). For the latter two components, we adopt $u_{\rm IR}=u_{\rm opt}=0.5\,\mathrm{eV\,cm^{-3}}$, as in the B0540$-$69 model of \citet{Tanaka2013}.} \red{The model fit yields a relatively weak outer-zone magnetic field, $B_{\rm out}=8.156\,\mu\mathrm{G}$, and a high cutoff energy for the inner-zone electron population, $E_{\rm in,c}=3.564\,\mathrm{PeV}$. The resulting magnetic-to-electron energy ratio is $W_B/W_e\sim\mathrm{few}\times10^{-4}$, indicating a particle-dominated configuration, as inferred for many other young PWNe \citep{2014JHEAp...1...31T}. The fitted SED and model parameters are presented in Figure~\ref{fig:b0540_two_zone_sed} and Table~\ref{tab:b0540_two_zone_parameters}, respectively. 
In this two-zone model, the inner-zone electron distribution of \src has an exponential cutoff at a few PeV, comparable in energy scale to the electrons required by leptonic interpretations of the PeV $\gamma$-rays detected from the Crab Nebula by LHAASO \citep{2021Sci...373..425L}. This model-dependent result suggests that \src may be capable of accelerating electrons to very high energies.}

% The \src PWN and the Crab Nebula are both luminous pulsar wind nebulae powered by young pulsars with high spin-down luminosities. Both exhibit complex spatial structure and broadband non-thermal emission. We therefore use the extensively observed Crab Nebula as a comparison source, showing its SED and a reference curve based on the constant-magnetic-field model of \citet{2010AA...523A...2M} in Figure~\ref{fig:b0540_two_zone_sed} alongside the \src results. 

\begin{table}[htb]
\centering
\caption{\red{Parameters of the two-zone model for \src PWN. }}
\label{tab:b0540_two_zone_parameters}
\begin{tabular}{lcc}
\hline\hline
\red{Parameter} & \red{Inner zone} & \red{Outer zone} \\
\hline
\red{$R$ / pc} & \red{0.1 (fixed)} & \red{1.8 (fixed)} \\
\red{$B$ / $\mu\mathrm{G}$} & \red{1010 (fixed)} & \red{8.156} \\
\red{$p_1$} & \red{1.88396} & \red{1.05396} \\
\red{$p_2$} & \red{3.07945} & \red{3.48819} \\
\red{$E_b$ / TeV} & \red{4.98772} & \red{0.434360} \\
\red{$E_c$ / PeV} & \red{3.564} & \red{1.000 (fixed)} \\
\hline
\end{tabular}
\end{table}

\section{Conclusion and Discussion}\label{sec:dis}

We have investigated the post-SRT evolution of PSR~\src using X-ray and $\gamma$-ray observations. In the X-ray band, the long-term flux history from 2000 to 2026 shows that the PSR$+$PWN luminosity increased by $\sim19\%$ after the SRT, corresponding to a $\sim26\%$ increase in the PWN component alone, and that this enhanced state has persisted for more than 14~yr. The pulsed X-ray luminosity remains unchanged across all epochs, and the photon index of the PSR$+$PWN system stays stable at $\Gamma \approx 1.8$--$1.9$. These results confirm that the X-ray brightening is produced by the non-pulsed component, i.e.\ the PWN, rather than by the pulsar itself. The absence of spectral evolution further suggests that the shape of the injected particle spectrum has not changed substantially; instead, the SRT most likely increased the normalization of the particle injection and/or the magnetic energy supplied to the PWN. In contrast, \textit{Fermi}/LAT shows no detectable long-term variation in either the pulsed or the non-pulsed flux between 2008 and 2025, although the current statistical uncertainties are large. %This decoupling between the X-ray brightening and the GeV stability rules out a simple broadband luminosity scaling and points to an energy-selective response of the PWN.

The updated timing solution confirms that \src remains in its high spin-down state. The braking index has continued to recover from $n \approx 0$ shortly after the SRT to $n \approx 1.53$ at present, but has not yet returned to the pre-SRT value of $\sim2.1$. This behavior is qualitatively consistent with the magnetic-axis-drift model \citep{2026ApJ...998..319B} and with the magnetospheric-reconfiguration scenario \citep{2020MNRAS.494.1865W}. The long recovery timescale ($>$14~yr) places constraints on the  processes within the neutron-star interior or magnetosphere.

% On the basis of the currently available information, this model provides a physically motivated interpretation consistent with the available SED and morphological evidence. Nevertheless, PSR B0540$-$69 is physically analogous to the Crab Nebula: both are luminous PWNe powered by young pulsars with high spin-down luminosities and exhibit complex, spatially dependent non-thermal emission \citep{2006ARAA..44...17G,2010AA...523A...2M,2023AA...671A..67D}. The present two-zone treatment is therefore necessarily oversimplified. A more complete description incorporating a continuous radial magnetic-field profile, advective and diffusive particle transport, time-dependent injection, radiative and adiabatic losses, and self-consistent energy-conservation equations would require substantially stronger observational constraints \citep{2009ApJ...703.2051G,2011ApJ...741...40T,2014JHEAp...1...31T}.

\red{Using the available multiwavelength observations, we have constructed 
the present broadband SED of the \src\ PWN, and built a simplified but effective two-zone model that separates the inner and outer emitting regions and provides a physically motivated interpretation consistent with the available SED and morphological evidence.  The model does not describe a continuous spatial variation in the magnetic field, particle transport, or time-dependent injection.} 
Nevertheless, PSR B0540$-$69 is physically analogous to the Crab: both have  luminous PWNe powered by young pulsars with high spin-down luminosities and exhibit complex, spatially dependent non-thermal emission \citep{2006ARAA..44...17G,2010AA...523A...2M,2023AA...671A..67D}. The present two-zone treatment is therefore necessarily oversimplified. A more complete description incorporating a continuous radial magnetic-field profile, advective and diffusive particle transport, time-dependent injection, radiative and adiabatic losses, and self-consistent energy-conservation equations would require substantially stronger observational constraints \citep{2009ApJ...703.2051G,2011ApJ...741...40T,2014JHEAp...1...31T}.
In particular, high-angular-resolution %X-ray 
observations are required to determine the energy-dependent source size, radial spectral softening, and surface-brightness profile of the nebula. Detailed phase-resolved spectroscopy from several hundred MeV to a few GeV would improve the separation of the pulsed magnetospheric and nebular components and test whether the inferred synchrotron-to-IC transition is genuine. Deeper multi-TeV observations would constrain the high-energy electron population and IC intensity in the outer zone, thereby testing whether electrons extending to several PeV are genuinely required to account for the broadband emission.

%Future high-angular-resolution observations that measure the energy-dependent source size, radial spectral softening, and surface-brightness profile would provide key tests of the adopted spatial division. Beyond the present-day SED, variability offers another way to probe particle supply and the evolution of the emitting regions.}

% \red{The Crab Nebula exhibits rapid and dramatic variability: its high-energy $\gamma$-ray flares show an additional synchrotron component lasting days. During the 2011 flare, the peak flux above $100,\mathrm{MeV}$ reached approximately 30 times its average level \citep{2011Sci...331..739A,2012ApJ...749...26B}. By comparison, the \src\ PWN has remained in an X-ray-bright state for more than 14~yr after the spin-down-rate transition (SRT), with no corresponding long-term variation detected in its GeV flux. The available data cannot establish whether this persistent X-ray brightening includes an additional high-energy synchrotron component analogous to that seen in the Crab flares, or whether the underlying microscopic mechanisms are the same. To examine the connection between activity of the central pulsar and changes in its nebula, we also consider the magnetar-like outbursts of PSR~J1846$-$0258 in Kes~75 and PSR~J1119$-$6127, together with the PWN changes observed afterward \citep{2008Sci...319.1802G,2008ApJ...686..508N,2018ApJ...856..133R,2016ApJ...829L..21A,2017ApJ...850L..18B}.}

\red{Besides the Crab, at the level of \emph{events}---a sudden change in the central pulsar followed by an X-ray response of its surroundings---the closer analogues in the literature are the 2006 magnetar-like outburst of PSR~J1846$-$0258 in Kes~75 \citep{2008Sci...319.1802G,2008ApJ...678L..43K,2011ApJ...730...66L} and the 2016 outburst of PSR~J1119$-$6127 \citep{2016ApJ...829L..21A,2017ApJ...850L..18B}. In both events, short hard X-ray bursts, large glitches, spectral hardening, and transient X-ray enhancements of the central pulsar (by factors of $\sim6$ and $\sim160$, respectively) developed and decayed on timescales of months to $\sim1$--$2$~yr, with radiated powers greatly exceeding any plausible change in the spin-down luminosity, requiring the release of internal magnetic free energy (i.e., magnetar-like activity) as the power source. In Kes~75, the extended X-ray brightening around the pulsar is small ($\sim10$--$20\%$), short-lived (months to $\sim1$~yr), and closely tied in time to the pulsar's burst episode; its interpretation as a genuine PWN response is not unambiguous, since contributions from a dust-scattering halo of the burst emission \citep{2010ApJ...710..227T} and from PSF wings of the strongly enhanced central point source \citep{2008ApJ...678L..43K,2008ApJ...686..508N} cannot be fully excluded. In PSR~J1119$-$6127 the associated PWN is too faint for any comparable measurement. PSR~\src\ is qualitatively different: no short bursts, no spectral hardening, and no glitch of comparable character have accompanied the SRT; the brightening is confined to the non-pulsed PWN component with an unchanged photon index; the excess PWN luminosity is well below the increase of the spin-down luminosity implied by the SRT; and the enhanced state has persisted for more than 14~yr. Rotational-energy re-partitioning therefore suffices to power the observed response, and there is no need to invoke internal magnetic free energy. In this sense, \src\ provides, to our knowledge, the clearest known example in which a persistent redistribution of a pulsar's rotational-energy loss can be tracked through the decade-scale radiative response of its nebula, complementing---rather than duplicating---the magnetar-like outburst events observed in PSR~J1846$-$0258 and PSR~J1119$-$6127.}

Overall, \src provides a rare case in which a sudden change in the spin-down power of a pulsar can be followed through the radiative response of its nebula. The persistent X-ray brightening, unchanged pulsed X-ray emission, stable GeV flux, and still-recovering braking index together indicate that the system has not yet returned to its pre-SRT equilibrium. Continued EP/FXT and multi-wavelength monitoring will be crucial for determining whether the elevated X-ray luminosity eventually decays as the braking index recovers, for detecting possible responses to future rotational instabilities, and for constraining how the additional spin-down power is partitioned between particles, magnetic field, and radiation within the PWN.

% --- Discussion ---

% \normalem
\begin{acknowledgements}

This work is supported by the National Natural Science Foundation of China (Grant Nos. 12673057, 12373051, 12033001, 12393851, 12393854, and 12473041), the China's Space Origins Exploration Program,  and the National Key R\&D Program of China No.\,2025YFF0511104. 
We thank Dr. Wynn Ho for kindly providing the ephemeris used in this work. We acknowledge the use of EP/FXT and the public data from the \emph{Fermi}/LAT and NICER data archives.
This work made use of data supplied by the UK Swift Science Data Centre at the University of Leicester.

The authors used ChatGPT/Claude to assist with language polishing and figure preparation. All content, code, and figures were reviewed and edited by the authors, who take full responsibility for the final manuscript.

\end{acknowledgements}

\appendix
\section{On- and Off-pulse Spectra made by EP/FXT}
\label{apx:onoffspc}

\begin{figure}[h]
    \centering
    \includegraphics[width=\linewidth]{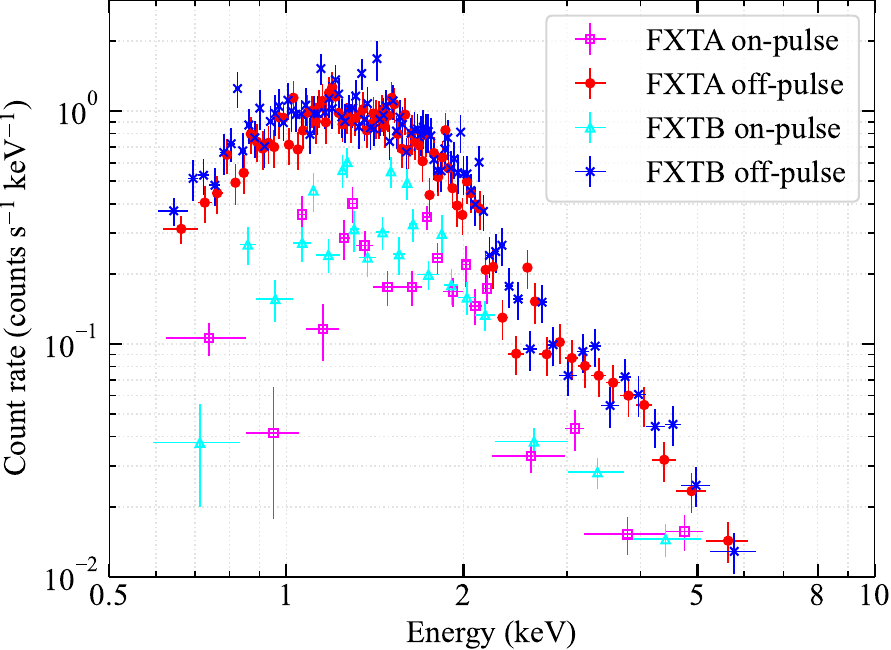}
    \caption{
    \red{Phase-resolved count-rate spectra of \src obtained with EP/FXT. The FXTA on- and off-pulse spectra are shown as magenta open squares and red filled circles, respectively, while the corresponding FXTB spectra are shown as cyan open triangles and blue crosses. The on-pulse spectra were extracted over the phase intervals $0$--$0.4$ and $0.6$--$1.0$, and the off-pulse spectra over phases $0.4$--$0.6$, as defined in Figure~\ref{fig:profile2}.}
    }
    \label{fig:fxt_on_off_spc}
\end{figure}

\renewcommand{\theequation}{\thesection.\arabic{equation}}
\renewcommand{\thefigure}{\thesection.\arabic{figure}}
\renewcommand{\thetable}{\thesection.\arabic{table}}

\makeatletter
\@addtoreset{equation}{section}
\@addtoreset{figure}{section}
\@addtoreset{table}{section}
\makeatother

\newpage

% \bibliographystyle{unsrt} 
% \bibliographystyle{unsrtnat}
% \bibliographystyle{apsrev4-1}
% \bibliography{psrb0540} 
%merlin.mbs apsrev4-1.bst 2010-07-25 4.21a (PWD, AO, DPC) hacked
%Control: key (0)
%Control: author (8) initials jnrlst
%Control: editor formatted (1) identically to author
%Control: production of article title (-1) disabled
%Control: page (0) single
%Control: year (1) truncated
%Control: production of eprint (0) enabled
%

\end{CJK*}

\end{document}